\documentclass[
letter,
twocolumn,
superscriptaddress,
amsmath,
amssymb,
prl,
nofootinbib
showkeys,
10pt,
floatfix,
nobibnotes,
aps,
fltpage
]{revtex4-2}

\usepackage{latexsym}
\usepackage{amsmath}
\usepackage{amssymb}
\usepackage[T1]{fontenc}
\usepackage[open]{bookmark}
\usepackage{hyperref}
\hypersetup{colorlinks=true,allcolors=blue}
\usepackage{upgreek}

\usepackage{orcidlink}

\usepackage{dcolumn}
\usepackage{bm}
\usepackage{xcolor}
\usepackage{hyperref}
\usepackage{cleveref}
\usepackage{float}
\usepackage{svg}
\usepackage{amssymb}
\usepackage{scalerel}
\usepackage{caption}
\usepackage{wasysym}
\usepackage{lipsum}
\usepackage{enumitem}

\providecommand{\pnl}[1]{{\textcolor{black}{(#1)}}}

\makeatletter
\pretocmd\frontmatter@keys@format{\addvspace{20\p@}}{}{}
\makeatother

\usepackage{svg}
\usepackage{cleveref}
\usepackage{siunitx}
\usepackage{caption}
\usepackage{multirow}
\usepackage{makecell}
\usepackage{tabularray}
\usepackage{makecell} 
\usepackage{array} 
\usepackage{graphicx} 
\usepackage{booktabs} 

\begin{document}
\title{Velocity-tunable exciton-photon hybridization in cathodoluminescence}

\author{Sven Ebel\,\orcidlink{0009-0005-3224-6413}}
\affiliation{POLIMA---Center for Polariton-driven Light--Matter Interactions, University of Southern Denmark, Campusvej 55, DK-5230 Odense M, Denmark}

\author{Martin Nørgaard\,\orcidlink{0009-0002-5412-4669}}
\affiliation{POLIMA---Center for Polariton-driven Light--Matter Interactions, University of Southern Denmark, Campusvej 55, DK-5230 Odense M, Denmark}

\author{Christian Nicolaisen Hansen\,\orcidlink{0009-0008-2350-7652}}
\affiliation{POLIMA---Center for Polariton-driven Light--Matter Interactions, University of Southern Denmark, Campusvej 55, DK-5230 Odense M, Denmark}

\author{N. Asger Mortensen\,\orcidlink{0000-0001-7936-6264}}
\affiliation{POLIMA---Center for Polariton-driven Light--Matter Interactions, University of Southern Denmark, Campusvej 55, DK-5230 Odense M, Denmark}
\affiliation{Danish Institute for Advanced Study, University of Southern Denmark, Campusvej 55, DK-5230 Odense M, Denmark}

\author{Sergii Morozov\,\orcidlink{0000-0002-5415-326X}}
\email{Corresponding author: semo@mci.sdu.dk}
\affiliation{POLIMA---Center for Polariton-driven Light--Matter Interactions, University of Southern Denmark, Campusvej 55, DK-5230 Odense M, Denmark}

\date{\today}

\begin{abstract}
\vspace{0.0cm}
\textbf{Abstract.} 
Exciton-photon hybridization is typically realised in geometrically defined optical cavities, where tunability is achieved by modifying either the cavity or the excitonic medium. Here, we investigate cathodoluminescence resonances arising from transition-radiation interference in suspended subwavelength films and explore their interaction with excitons in transition metal dichalcogenides. We demonstrate, both experimentally and using an analytical transition-radiation model, that these resonances hybridize with excitonic transitions and can be tuned continuously by varying the electron energy. Over experimentally accessible acceleration voltages of 10--30\,keV, this tuning is sufficient to shift the resonances through the excitonic transitions of WS$_2$ thin films. The resulting tuning depends on both film thickness and electron velocity, establishing the latter as an external and continuous knob for exciton--photon coupling. This approach enables tunable hybridization without structural modification and provides a free-electron-driven nanoscale platform for studying exciton--light interactions.
\vspace{0.3cm}
\end{abstract}

\maketitle

\section{Introduction}
Hybridization between excitons and confined optical modes provides a powerful route to engineer mixed light–matter states with tailored spectral and dynamical properties~\cite{Schneider2018}. 
A broad range of resonant architectures has been developed to realise such interactions, including planar Fabry--P{\'e}rot microcavities~\cite{Weisbuch1992}, photonic crystals~\cite{Zhang2018}, plasmonic resonators~\cite{Sugawara2006,Liu2016}, Mie resonators~\cite{Todisco2020,Wang2020}, and bound states in the continuum~\cite{Maggiolini2023,Weber2023}. 
These platforms provide field confinement and spectral selectivity across multiple length scales, enabling hybridization with diverse excitonic media.
Tunability in these systems is commonly achieved by modifying either the photonic resonance or the excitonic medium. The photonic resonance can be controlled structurally, for example by changing the thickness or lateral extent of Fabry--P{\'e}rot resonators~\cite{Dufferwiel2015,Urbonas2016,Lackner2021} or by tailoring the dimensions and arrangement of nanoscale resonators~\cite{VuCam2025}. Alternatively, the excitonic response can be tuned through electrical gating~\cite{Sidler2016,Munkhbat2020,Hoekstra2026} or chemical control~\cite{Ahn2020,Friedrich2025}. A further widely used approach is angle tuning, where changing the optical incidence or collection angle modifies the sampled in-plane photon momentum and thereby shifts the observed photonic resonance relative to the exciton. This strategy is commonly employed in planar Fabry--P{\'e}rot microcavities and diffractive surface-lattice-resonance systems~\cite{Weisbuch1992,Rodriguez2013,Guo2017}.

More generally, hybridization does not rely on a specific cavity design, but rather on the presence of a well-defined optical eigenmode. Such modes may be supported by an external resonator or arise intrinsically from the geometry and dielectric contrast of the structure, giving rise to so-called self-hybridized exciton–photon systems~\cite{Canales2021,vanVugt2011,Verre2019,Tserkezis2024}. The van~der~Waals (vdW) semiconductors are particularly well suited in this regime, as their large exciton oscillator strengths and strong dielectric contrast support intrinsic optical modes that readily couple to excitonic resonances~\cite{Wang2016,Munkhbat2018,Dirnberger2023,Black2024,Anantharaman2024,Ziegler2025}. In such self-hybridized systems, angle- or momentum-resolved optical spectroscopy can be used to access the intrinsic polariton dispersion~\cite{Munkhbat2018,Dirnberger2023}. However, the optical modes themselves remain largely fixed by the geometry and dielectric response of the structure, which limits the scope for active tuning once the sample is fabricated.

While optical spectroscopy has been central to establishing cavity and self-hybridized platforms, accessing hybridization at the nanoscale often requires probes that combine broadband excitation with nanometer spatial resolution. Free-electron-based spectroscopies meet these requirements and have therefore emerged as powerful tools for investigating exciton-photon hybrid systems in real space~\cite{Kociak_2025,Song2021,Yankovich2019,Bitton2020,Chahshouri2022,Nerl2024}. Cathodoluminescence (CL) microscopy provides high spatial and spectral resolution and has been used to map plasmon--exciton hybridization~\cite{Davoodi2022} as well as the impact of geometric confinement on exciton--light hybridization~\cite{Taleb2021}.

Here we investigate transition radiation (TR), the emission generated when an energetic free electron crosses an interface between two media~\cite{Ginzburg1996,Ritchie1962}, in the presence of excitons in transition metal dichalcogenide (TMD) thin films. In thin films, TR emitted at the two interfaces can interfere constructively, giving rise to  free-electron-defined resonances~\cite{Arakawa1964,Yamamoto1996,Lin2018,Fiedler2022}, whose spectral position is tunable via the electron velocity.  We show that TR resonances in subwavelength dispersive films hybridize with the excitonic response at electron energies typical of scanning electron microscopy. 
The resulting spectral response depends on the incident electron energy, enabling velocity--tunable exciton--photon hybridization without any geometric modification of the cavity. 
This free-electron-tunable platform therefore provides a route to tunable exciton–photon hybridization.

\section{Results}
\subsection{Transition radiation resonances}

\begin{figure}[t!]
 \includegraphics[width=0.99\linewidth]{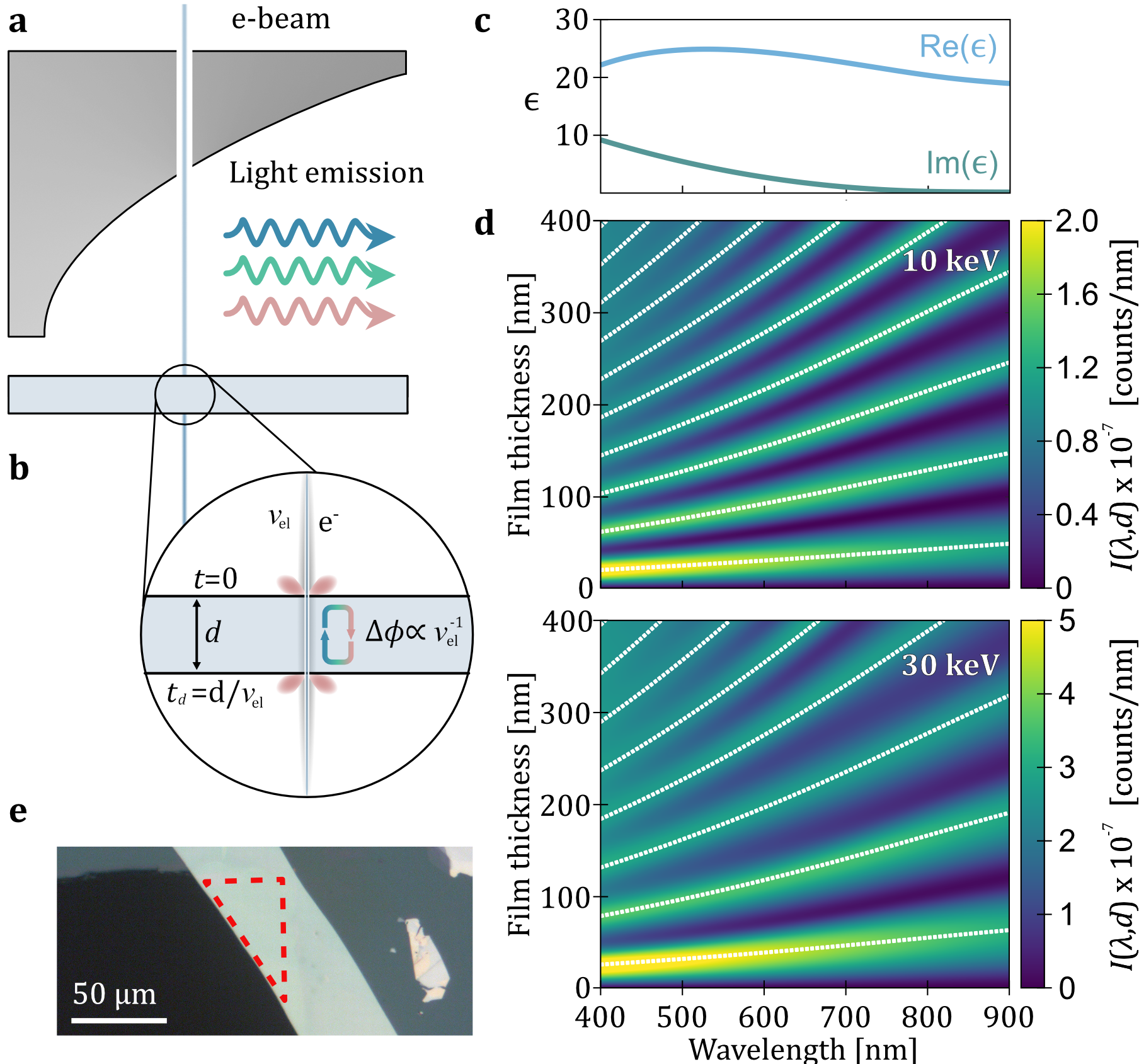}
\caption{\textbf{Transition radiation resonances.}
\pnl{a} Experimental setup showing a parabolic mirror positioned in the electron beam path to collect the generated transition radiation (TR) and direct it to a spectrometer.
\pnl{b} Schematic illustration of electron-velocity-dependent resonant TR in thin films of thickness $d$.
\pnl{c} Dielectric function of a model semiconductor with a bandgap of 1.5\,eV.
\pnl{d} Calculated TR intensity for dielectric films described by the dielectric response shown in panel~\pnl{c}, plotted as a function of film thickness $d$ and emission wavelength for electron energies of 10 and 30\,keV.
\pnl{e}~Optical image of a suspended 83\,nm thick WS$_2$ crystal. The red dashed line marks the suspended region.}
\label{fig-intro}
\end{figure}

We consider a free electron with velocity $v_{\mathrm{el}}$ normally traversing a thin film of thickness $d$ and wavelength-dependent complex-valued isotropic dielectric function $\epsilon(\lambda)$, freely suspended in vacuum [Fig.~\ref{fig-intro}\pnl{a}]. An electron interacting with such a system generates TR at each film interface as well as Cherenkov radiation within the film. An analytical treatment of this problem was developed by Ritchie and Eldridge~\cite{Ritchie1962}. 
For normal electron-beam incidence, the intensity per incident electron, emitted into a unit solid angle at polar angle $\theta$ and per unit wavelength $\lambda$ in the backward direction relative to the electron trajectory, is given by:
\begin{equation}\label{TR_model}
    I(\theta,\lambda,d,\beta)=\frac{\alpha\beta^2}{\pi^2\lambda}\mu^2(1-\mu^2)\left|\frac{\gamma}{\Delta}\right|^2
\end{equation}
with
\begin{equation}\label{Term_contr.}
\begin{aligned}
\gamma = 
&\left[\frac{\epsilon - \beta\sigma}{1 - \mu^2\beta^2}- \frac{1}{1 + \beta\sigma}
\right](\mu\epsilon + \sigma)\, e^{-i t \sigma}
\\[4pt]
&+
\left[-\frac{\epsilon + \beta\sigma}{1 - \mu^2\beta^2}+ \frac{1}{1 - \beta\sigma}\right](\mu\epsilon - \sigma)\, e^{i t \sigma}
\\[4pt]
&-2\sigma\left[\frac{\epsilon}{1+\beta\mu}-\frac{1-\beta\epsilon\mu}{1-\beta^2\sigma^2}\right]\, e^{i t/\beta}\,,
\end{aligned}
\end{equation}
and $\sigma=(\epsilon-1+\mu^2)^{\frac{1}{2}}$, $\beta=v_{\mathrm{el}}/c$, $\mu=\cos\theta$, $t=2\pi d/\lambda$, $\Delta=(\mu\epsilon-\sigma)^2e^{it\sigma}-(\mu\epsilon+\sigma)^2e^{-it\sigma}$, $c$ the speed of light, and $\alpha$ the fine structure constant. 
Equation~\eqref{Term_contr.} contains both Cherenkov-radiation and transition-radiation contributions. Following Yamamoto et al.~\cite{Yamamoto1996}, the resonance condition can be interpreted in the limit where the Cherenkov contribution is weak, such that the signal is dominated by interference of transition radiation emitted at the two film interfaces. This limit describes the resonance positions in the present thin-films, as discussed in Supplementary Note~S1, while the full expression is retained for all calculated spectra. Constructive interference between those terms occurs when the phase condition $\phi=t\sigma + \frac{t}{\beta} = 2\pi\left(m-\frac{1}{2}\right)$ is satisfied, where $m$ is an integer. This condition is illustrated in Fig.~\ref{fig-intro}\pnl{b} and shows that the phase delay $\Delta\phi$ between successive TR excitation events is governed by the film thickness and the electron velocity. From this phase condition, we derive an expression for the resonance energy:

\begin{equation}\label{Dispersion}
E_{\mathrm{ph}}=\frac{hv_{\mathrm{el}}}{d}\left(m-\tfrac{1}{2}\right)\left(1+\frac{v_{\mathrm{el}}}{c}\sqrt{\epsilon-\sin^2\theta}\,\right)^{-1}.
\end{equation}

We next evaluate Eq.~\eqref{TR_model} for a dispersive dielectric film described by a model dielectric function representative of a generic semiconductor [see Eq.~\eqref{semiconductor} in Methods and Fig.~\ref{fig-intro}\pnl{c}], as a function of film thickness and electron energy. The results, shown in Fig.~\ref{fig-intro}\pnl{d}, reveal that the TR spectrum exhibits thickness-dependent resonances reminiscent of Fabry--P{\'e}rot cavity modes, characterized by standing-wave interference that produces spectral maxima and minima. In contrast to a conventional Fabry--P{\'e}rot cavity, the number of resonances exhibits a pronounced dependence on the electron velocity, with more resonances appearing over the same film-thickness range for lower electron velocities.
This behaviour of the TR resonance is well captured by Eq.~\eqref{Dispersion}, which accurately reproduces both the thickness and electron-energy dependence of the constructive-interference resonances, as indicated by the white overlays in Fig.~\ref{fig-intro}\pnl{d}. The corresponding energy tunability is discussed in more detail in Supplementary Note~S1.

\subsection{Probing transition radiation resonances}
To experimentally validate the TR model and examine its interaction with excitonic transitions, we use vdW materials. This material class offers both dispersive systems without additional sharp excitonic features and semiconducting compounds with pronounced excitonic resonances, while enabling straightforward fabrication and thickness control in suspended thin films. The experimental setup, schematically shown in Fig.~\ref{fig-intro}\pnl{a}, is based on a scanning-electron microscope (SEM) operating in a configuration with a Schottky field-emission beam and equipped with a parabolic mirror positioned above the sample stage. The mirror features a central aperture for free passage of the electron beam and collects the generated CL, which is subsequently directed to the optical detection system. As illustrated in Fig.~\ref{fig-intro}\pnl{e}, selected vdW crystals were transferred onto transmission electron microscopy (TEM) windows such that the crystals are freely suspended, providing two parallel vacuum-vdW material interfaces during the measurements. The thickness of the selected crystals was determined by optical reflectivity measurements, allowing us to determine the thickness with an error margin of $\pm 2$\,nm (see Methods and Supplementary Note~S2 for more details). This deterministic fabrication procedure enables efficient selection of crystals with the desired thickness and lateral size.
\begin{figure}
\centering
\includegraphics[width=0.9\linewidth]{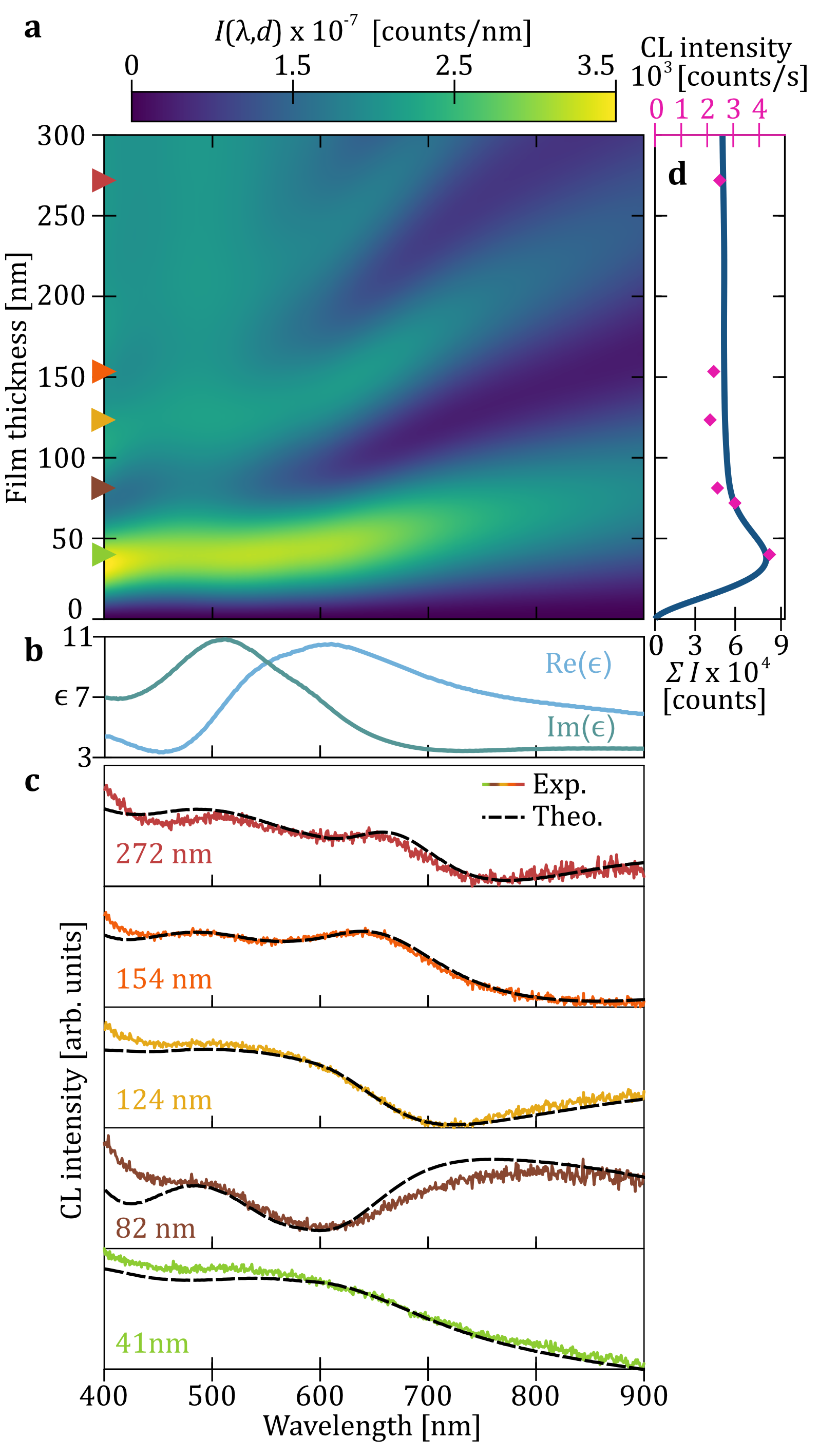}
\caption{\textbf{Transition radiation from NbSe$_2$ films.} 
\pnl{a}~Calculated TR intensity from NbSe$_2$ films as a function of film thickness $d$ and emission wavelength for a free-electron energy of 30\,keV.
\pnl{b} Experimental in-plane dielectric function of NbSe$_2$ as reported by Munkhbat \emph{et~al.}~\cite{Munkhbat2022}.
\pnl{c} Experimental CL spectra at 30\,keV and corresponding theoretical TR spectra for NbSe$_2$ films with thicknesses of 272, 154, 124, 82, and 41\,nm. The selected thicknesses are indicated by triangles in panel \pnl{a}. 
\pnl{d} Spectrally integrated experimental CL (pink diamonds) and theoretical TR (blue) intensities. In the experiments, the beam current was kept at 1.4\,nA and the exposure time was 90\,s. An extended thickness series including additional NbSe$_2$ flakes is provided in Supplementary Note~S3.}
\label{NbSe2}
\end{figure}

 We first investigate niobium diselenide (NbSe$_2$) as a dispersive reference system to test the TR model of Eq.~\eqref{TR_model}. NbSe$_2$ is a semimetal characterized by a strongly dispersive dielectric function, exhibiting low optical losses towards the infrared spectral region and increased losses in the blue spectral region, the latter attributed to absorption channels arising from intraband transitions~\cite{Hill2018,Munkhbat2022}. Previous CL studies have shown that the bulk CL response of NbSe$_2$ is dominated by TR, allowing contributions from other CL generation mechanisms to be neglected in the present analysis~\cite{Ebel2025}. 
Figure~\ref{NbSe2}\pnl{a} shows the TR intensity calculated from Eq.~\eqref{TR_model} for NbSe$_2$ films as a function of film thickness at an electron energy of 30\,keV. The in-plane dielectric function of NbSe$_2$ used in the calculations [Fig.~\ref{NbSe2}\pnl{b}] was obtained from spectroscopic ellipsometry measurements reported by Munkhbat \emph{et~al.}~\cite{Munkhbat2022}. The calculations in Fig.~\ref{NbSe2}\pnl{a} reveal that the TR spectrum exhibits the expected resonant behaviour in the low-loss spectral region, forming distinct thickness-dependent interference maxima and minima. In contrast, the high-loss spectral region is characterized by a reduced interference contrast, indicating the suppression of interference due to increased optical losses. 

We experimentally probe this behaviour by measuring a series of suspended NbSe$_2$ crystals with different thicknesses (indicated by coloured triangles in Fig.~\ref{NbSe2}\pnl{a}).
Fig.~\ref{NbSe2}\pnl{c} compares the calculated and measured CL spectra for crystals with thicknesses of 274, 154, 124, 82, and 41\,nm.
Each measured CL spectrum is plotted alongside the corresponding calculated TR spectrum, demonstrating good agreement and capturing the characteristic interference features of the TR response. All experimental spectra are normalized and aligned to the theoretical TR spectrum at 600\,nm, corresponding to the spectral region of highest sensitivity of our detection system. Consequently, the largest deviations between experimental and theoretical spectra are observed towards the blue spectral region, which coincides with the lowest detection sensitivity and with the general increase of the TR intensity scaling as $1/\lambda$. Another source of deviation arises from the finite out-of-plane dielectric response of vdW materials~\cite{Nrgaard2025}, which is not captured by the isotropic model considered here.

Finally, Fig.~\ref{NbSe2}\pnl{d} shows the spectrally integrated TR intensity as a function of film thickness, overlaid with the corresponding spectrally integrated CL intensity, again demonstrating good agreement between the theoretical TR response and the experimental measurements. Notably, the intensity exhibits an approximately twofold enhancement for few-nanometer-thick films, corresponding to the first-order interference maximum. This behaviour arises from the lossy nature of NbSe$_2$, which leads to damping of higher-order resonances, while the first-order resonance remains weakly dispersive at small thicknesses.

\subsection{Hybridization with excitons}

\begin{figure}
\centering
\includegraphics[width=0.9\linewidth]{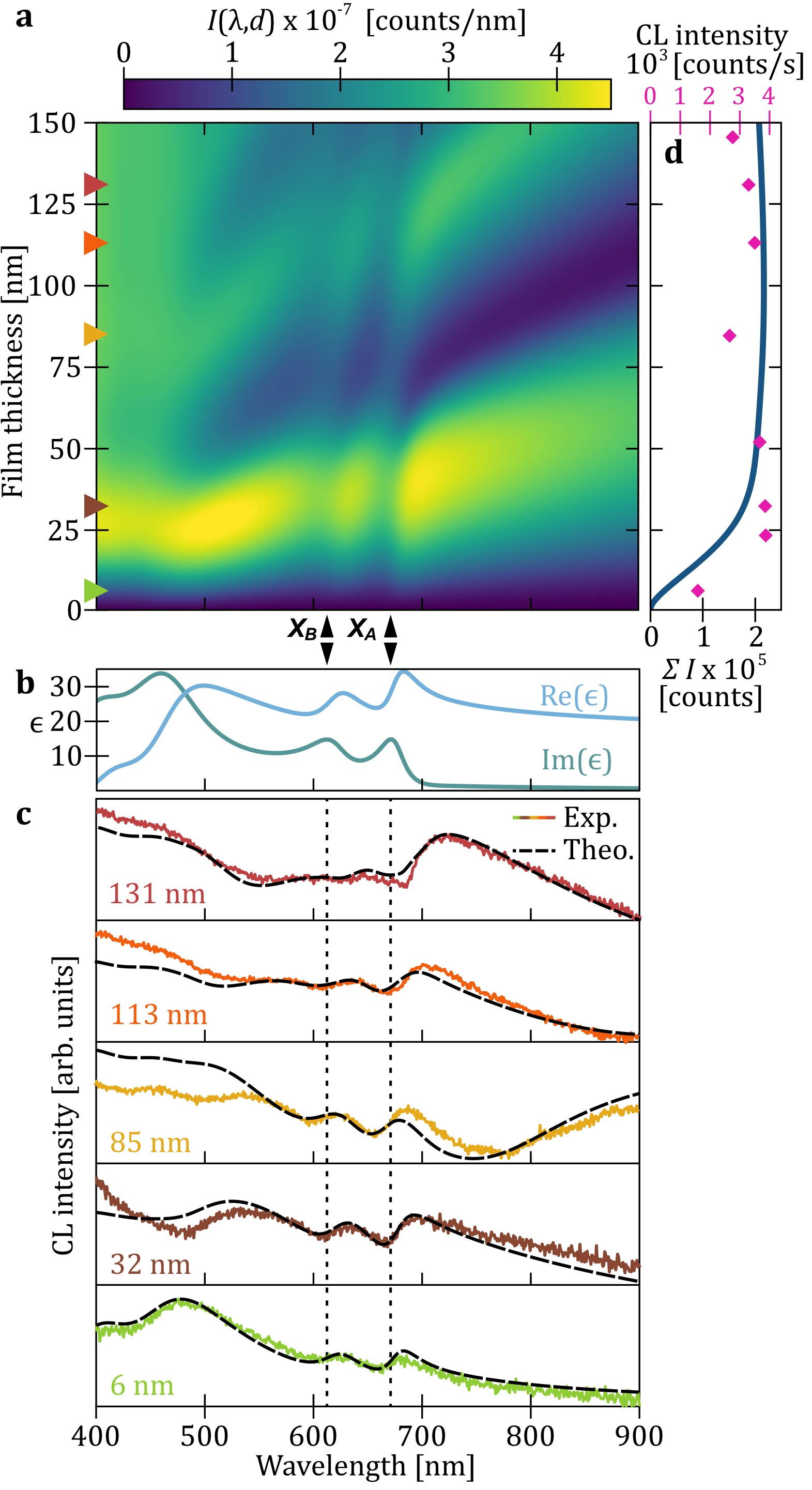}
\caption{\textbf{Transition radiation from MoS$_2$ films.} \pnl{a}~Calculated TR intensity from MoS$_2$ films as a function of film thickness $d$ and emission wavelength for a free-electron energy of 30\,keV.
\pnl{b} Experimental in-plane dielectric function of MoS$_2$ as reported by Ermolaev \emph{et~al.}~\cite{Ermolaev2021}.
\pnl{c}~Experimental CL spectra at 30\,keV and corresponding theoretical TR spectra for MoS$_2$ films with thicknesses of 131, 113, 85, 32, and 6\,nm. The selected thicknesses are indicated by triangles in \pnl{a}. 
\pnl{d} Spectrally integrated experimental CL (pink diamonds) and theoretical TR (blue) intensities. In the experiments, the beam current was kept at 1.4\,nA and the exposure time was 90\,s. An extended thickness series including additional MoS$_2$ flakes is provided in Supplementary Note~S3.}
\label{MoS2}
\end{figure}

Having established that Eq.~\eqref{TR_model} accurately describes the CL response of subwavelength dispersive films, we now examine how excitons modify the TR response. 
As a model system, we investigate molybdenum disulfide (MoS$_2$), a semiconducting vdW material whose bulk CL response exhibits pronounced excitonic transitions~\cite{Ebel2025}. 
MoS$_2$ features A and B excitons at 672\,nm and 612\,nm, respectively, as extracted from the in-plane dielectric function measured by spectroscopic ellipsometry by Ermolaev \emph{et~al.}~\cite{Ermolaev2021} and shown in Fig.~\ref{MoS2}\pnl{b}.
Fig.~\ref{MoS2}\pnl{a} shows the calculated TR spectrum of MoS$_2$ films as a function of film thickness. As in the case of NbSe$_2$, thickness-dependent resonant features characteristic of the TR response of a dispersive film are observed. 
In contrast to NbSe$_2$, however, the TR spectrum of MoS$_2$ exhibits pronounced spectral dips within the resonance structure that coincide with the excitonic resonances.

Experimentally, we probe this behaviour by measuring the CL response of suspended MoS$_2$ crystals with thicknesses selected to sample on- and off-resonant TR conditions.
Fig.~\ref{MoS2}\pnl{c} shows the CL spectra of MoS$_2$ crystals with thicknesses of 6, 32, 85, 113, and 131\,nm. 
The 32\,nm and 113\,nm crystals correspond to the expected maxima of the first- and second-order TR resonances, respectively. 
For all thicknesses, the measured CL spectra agree well with the analytically calculated TR response, reproducing both the thickness-dependent resonances and the modifications induced by the A and B excitons.
Similar exciton-related dips with adjacent maxima have previously been reported in CL studies of thin WSe$_2$ crystals by Taleb \emph{et~al.}~\cite{Taleb2021}, where they were attributed to hybridization between excitonic resonances and the waveguide modes of the crystal.
To emphasize the difference between the here reported TR resonances and thin film unique waveguide modes it is important to consider that waveguide modes are uniquely determined by the optical eigenvalue problem of the structure, meaning that their existence and properties remain independent from the possible source of excitation. This implies that a change of electron velocity does not tune the waveguide-mode dispersion itself; it only changes the momentum and frequency components with which the electron can excite a given mode, and therefore modifies the excitation probability only~\cite{DiGiulio2024}.

A closer analysis of the CL spectra in Fig.~\ref{MoS2}\pnl{c} reveals that the spectral positions of the features surrounding the exciton energy depend on the crystal thickness. 
For the resonant crystals (32 and 113\,nm), the spectral dips coincide with the exciton energy, whereas the off-resonant crystals (6, 85, and 131\,nm) exhibit spectrally shifted dips. This behaviour indicates that matching or detuning the TR resonance relative to the exciton energy modifies the excitonic contribution to the CL spectrum, consistent with hybridization between the TR resonance and the exciton.

Analogous to the analysis for NbSe$_2$, Fig.~\ref{MoS2}\pnl{d} shows the spectrally integrated TR intensity as a function of film thickness, overlaid with the corresponding CL intensity. The measured spectrally integrated CL intensities agree well with theory and confirm that the first-order resonance enhances the CL response for MoS$_2$ crystals with thicknesses around 32\,nm.

\subsection{Electron-energy-controlled detuning}

\begin{figure}
    \centering    \includegraphics[width=0.95\linewidth]{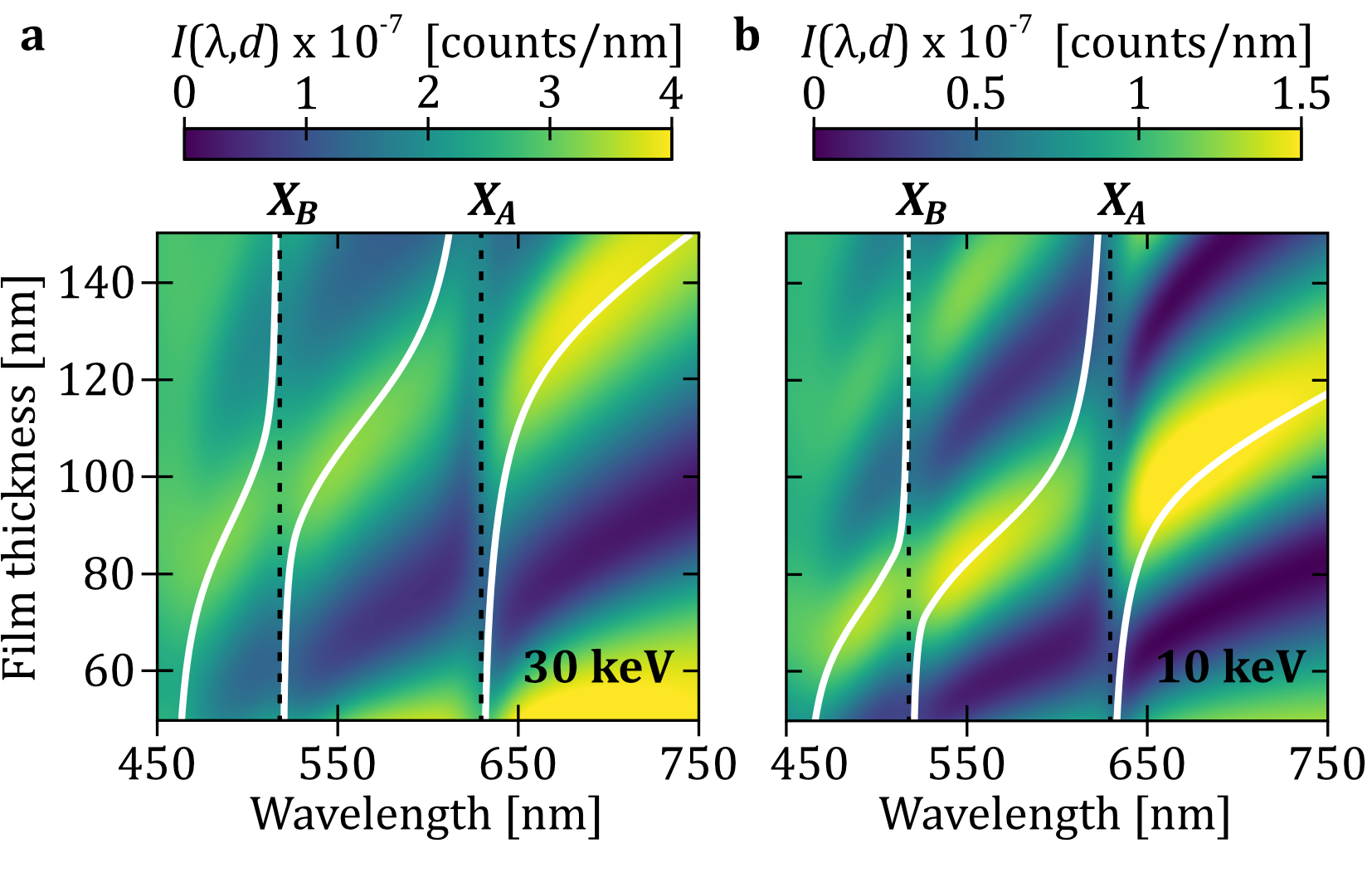}
    \caption{\textbf{Electron-energy dependent hybridization of TR resonances and excitons in WS$_2$ films}. Calculated TR intensity spectra from WS$_2$ films as a function of film thickness $d$ and emission wavelength for free-electron energies of \pnl{a} 30\,keV and \pnl{b} 10\,keV. Black dashed lines indicate the spectral positions of the A and B excitons in WS$_2$. White solid lines show coupled-oscillator fits to the calculated TR resonance maxima.
}
    \label{TunableTheo}
\end{figure}

To further investigate the tunability of hybridization between TR resonances and excitons, we turn to thin tungsten disulfide (WS$_2$) crystals.
Compared to MoS$_2$, WS$_2$ exhibits stronger excitonic oscillator strengths and more spectrally isolated A and B excitons, enabling a clearer spectral separation of the hybridized features. 
WS$_2$ features A and B excitons at 630\,nm and 518\,nm, respectively, as extracted from the in-plane dielectric function measured by Munkhbat \emph{et~al.}~\cite{Munkhbat2018}.
Fig.~\ref{TunableTheo} shows the calculated TR spectrum of WS$_2$ films as a function of thickness for electron energies of 10 and 30\,keV, with the exciton positions indicated by black dashed lines. 
Here we investigate WS$_2$ films in the thickness range of 50--150\,nm, which corresponds to the regime of the second-order TR resonance ($m=2$). 
For both electron energies, the TR resonance exhibits a pronounced bending of the resonance maximum near the exciton energies, consistent with the formation of hybridized light–matter states. 
This behaviour is reproduced by coupled-oscillator calculations (see Methods), in which the resonance energy obtained from Eq.~\eqref{Dispersion} is coupled to the excitonic transitions of WS$_2$.

\begin{figure*}
    \centering   \includegraphics[width=0.9\linewidth]{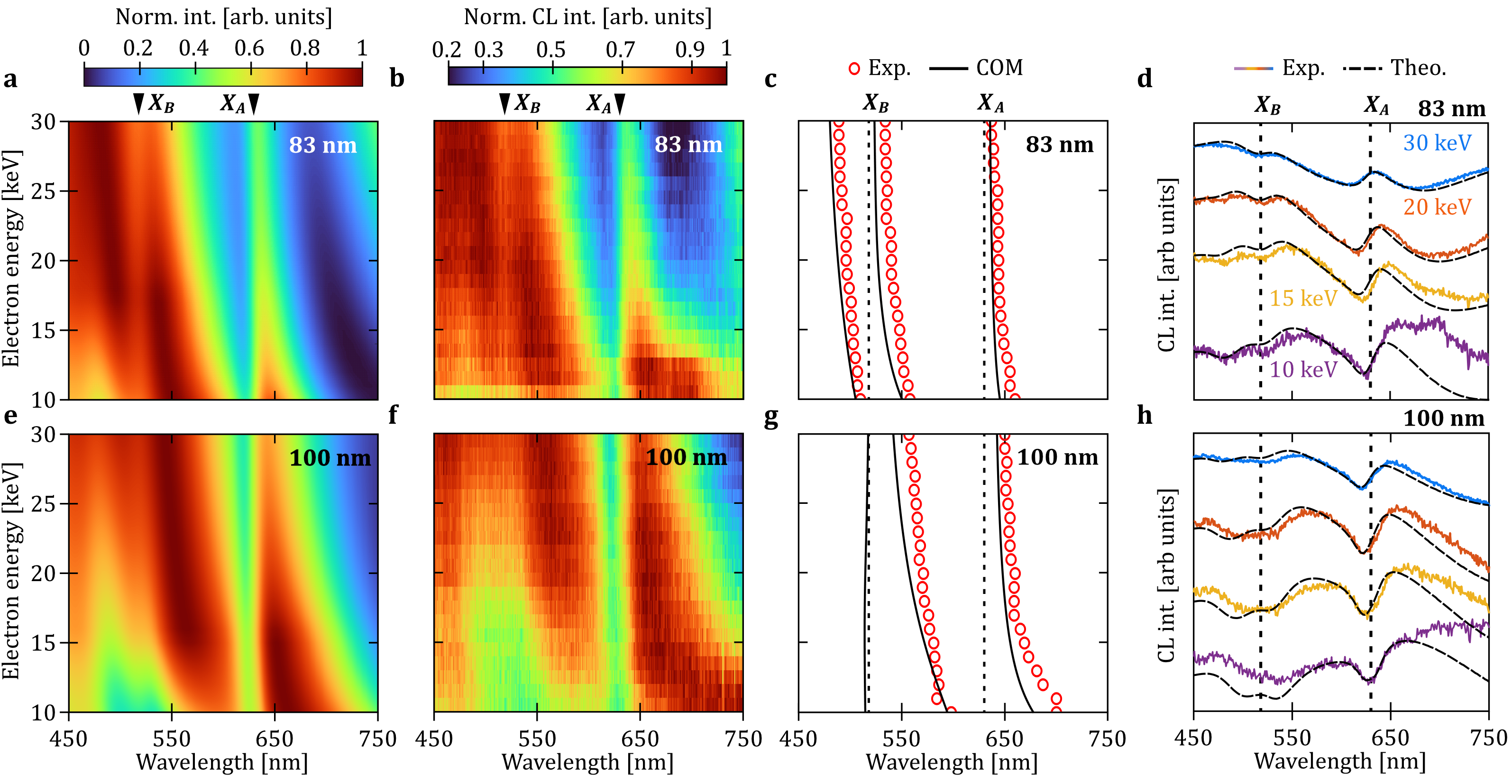}
    \caption{\textbf{Tunable hybridization between TR resonances and excitons in WS$_2$ films.} 
    \pnl{a,e} Calculated electron-energy-dependent TR spectra for WS$_2$ films with thicknesses of 83\,nm \pnl{a} and 100\,nm \pnl{e}. \pnl{b,f} Measured electron-energy-dependent CL spectra for WS$_2$ films with thicknesses of 83\,nm \pnl{b} and 100\,nm \pnl{f}. 
    \pnl{c,g} Extracted experimental spectral positions of the intensity maxima in the CL spectra (red circles). 
    Black solid lines show solutions of the coupled-oscillator model (COM) illustrating the tunability of the TR–exciton hybridization with electron energy. 
    \pnl{d,h} Experimental CL spectra from WS$_2$ films with thicknesses of 83\,nm \pnl{d} and 100\,nm \pnl{h} acquired at electron energies of 30, 20, 15, and 10\,keV. The spectra are normalized around 600\,nm and compared to calculated TR spectra of the same thickness and electron energy. During the measurements, the beam current was maintained at 1.4\,nA, with exposure times of 90--120\,s.}
    \label{TunableExp}
\end{figure*}
Changing the electron energy shifts the TR resonance according to Eq.~\eqref{Dispersion}, thereby modifying its dispersion. Consequently, the bending of the resonance maximum occurs at different film thicknesses, reflecting the tunability of the TR resonance relative to the exciton energies. Within the coupled-oscillator framework, this behaviour appears as a shift of the hybridized features with changing electron energy. Together, these results establish electron energy as a continuous and controllable parameter for tuning the TR resonance across the excitonic transitions.

We first probe electron-energy control of the hybridized response in an 83\,nm thick WS$_2$ crystal. The calculated TR spectrum [Fig.~\ref{TunableExp}\pnl{a}] shows that varying the electron energy tunes the TR resonance through the A-exciton energy. Consistent with this prediction, CL spectra acquired between 10 and 30\,keV (in 1\,keV steps) reproduce the same dispersion trend [Fig.~\ref{TunableExp}\pnl{b}].
Representative spectra in Fig.~\ref{TunableExp}\pnl{d} illustrate the evolution from a detuned, Lorentzian-like excitonic peak at 30\,keV to an asymmetric line shape at intermediate energies (20 and 15\,keV), and finally to a pronounced dip accompanied by two adjacent maxima when the TR resonance overlaps spectrally with the A exciton at 10\,keV. 
The largest deviation from the calculated TR response occurs at 10\,keV between 650 and 750\,nm, where the measured intensity exceeds the prediction, likely due to additional incoherent contributions that become more prominent at low electron energies (Supplementary Note~S4).
A distinct behaviour is observed near the B exciton, where the spectra exhibit a dip fixed at the exciton energy with two adjacent maxima that shift to lower energies as the acceleration voltage decreases. 
Concurrently, the higher-energy maximum decreases in amplitude while the lower-energy maximum increases, consistent with tuning the TR resonance through the B-exciton energy. 
To quantify this behaviour, we extract the energies of the maxima from the full data set [Fig.~\ref{TunableExp}\pnl{c}]. 
A coupled-oscillator model reproduces the measured dispersion using coupling strengths of 250\,meV (TR resonance--A exciton) and 200\,meV (TR resonance--B exciton), confirming that the evolving maxima originate from hybridized states formed by the interaction of the TR resonance with the excitonic transitions in WS$_2$.
These values are comparable to coupling strengths reported for exciton-photon self-hybridization in TMD thin films, which typically lie in the 100-300\,meV range, with larger splittings reported for WS$_2$ films~\cite{Wang2016,Munkhbat2018,Taleb2021}.

For a thicker 100\,nm WS$_2$ crystal, the TR resonance can be tuned across the A-exciton energy within the accessible electron-energy range [Fig.~\ref{TunableExp}\pnl{e}]. 
The corresponding CL spectra acquired between 10 and 30\,keV (in 1\,keV steps) reproduce the tuning trend [Fig.~\ref{TunableExp}\pnl{f}], and representative spectra are shown in Fig.~\ref{TunableExp}\pnl{h}. 
In contrast to the 83\,nm crystal, the A-exciton region is already characterised by a dip-like feature at high electron energies, accompanied by two adjacent maxima.
As the electron energy is reduced, the dip shifts toward the A-exciton energy and becomes centred at 10\,keV, while both maxima redshift, with the lower-energy maximum exhibiting the stronger shift. 
To quantify this evolution, we again extract the energies of the maxima and compare them to the coupled-oscillator model. 
The extracted dispersion shows good agreement with the model and reveals an increasingly asymmetric redshift below $\sim$20\,keV, consistent with tuning through the excitonic resonance. 
Around the B exciton, the response corresponds to the inverse detuning scenario compared with the A-exciton behaviour in the 83\,nm crystal, and a predominantly Lorentzian-like peak is recovered at the B-exciton energy at intermediate electron energies [Fig.~\ref{TunableExp}\pnl{h}]. 

The electron-energy tunability of TR resonances depends on the resonance order $m$. While the 83\,nm and 100\,nm crystals probe the second-order resonance ($m=2$), thinner films enter the first-order regime ($m=1$), for which Eq.~\eqref{Dispersion} predicts a reduced energy tunability due to the $(m-1/2)$ dependence. Consequently, the $m=1$ resonance exhibits a comparatively flat dispersion and, in turn, only weak tuning of the hybridized spectral features can be expected. This behaviour is confirmed by additional measurements on a 55\,nm thick flake, presented in Supplementary Note~S5.

\section{Discussion}
In this work, we have demonstrated that TR resonances supported by suspended subwavelength dispersive films provide a versatile platform for exciton--photon hybridization under free-electron excitation. 
We show that the spectral response of TR can be tuned continuously through excitonic transitions by varying either the film thickness or the electron energy. This enables controlled detuning between the photonic resonance and excitons in TMD thin films, resulting in spectral features consistent with hybridized light--matter states.
The extracted TR--exciton coupling strengths are comparable to values reported for self-hybridized systems~\cite{Wang2016,Munkhbat2018,Taleb2021}. In WS$_2$, the combination of film-thickness selection and electron-energy tuning allows both A- and B-exciton interactions to be analysed within the same TR-resonance framework. This is particularly useful for the B exciton, for which explicit experimental coupling estimates are less frequently reported than for the A exciton, and highlights the broadband spectral access provided by CL.

In contrast to conventional cavity platforms, where tunability is typically achieved through geometric modification of the resonator or active control of the excitonic medium, the present approach introduces the electron energy as an external and continuous tuning parameter. The TR interference thus acts as an electron-defined optical resonance whose dispersion can be engineered without altering the physical structure of the film. This capability is particularly relevant for self-hybridized systems in vdW semiconductors, where the intrinsic optical mode is otherwise fixed by geometry.

More broadly, these results establish TR resonances under electron-beam excitation as a platform for studying tunable hybrid light-matter interactions. The ability to manipulate the photonic resonance via electron velocity, while simultaneously leveraging the broadband character and nanometer spatial resolution of cathodoluminescence, opens new opportunities for probing exciton-photon coupling in engineered self-hybridized photonic systems. In contrast to optical excitation schemes that may require different light sources or angle-resolved configurations to access different spectral regions, free-electron excitation provides broadband access to optical responses from the ultraviolet (UV) to the infrared (IR) in a single experimental geometry.
This capability may be particularly useful for complex vdW devices, where local gating, electrostatic doping, or strain can generate spatially varying excitonic and photonic environments on length scales below the optical diffraction limit~\cite{Sidler2016,Munkhbat2020,Ai2025}. In such systems, electron-energy tuning would allow the detuning dependence of hybrid modes to be probed locally on the same device region, without modifying the structure or changing the optical excitation geometry. Beyond self-hybridized vdW semiconductors, this approach may therefore be extended to cavities and optical resonances formed in vdW heterostructures, where electron-controlled tuning could provide additional flexibility for investigating exciton-light interactions in complex layered architectures~\cite{Frg2019,Moore2025}. In this context, electron-controlled tuning may also be useful for studying hybridization involving spatially localized excitonic states, including moir\'e excitons in twisted or lattice-mismatched heterostructures~\cite{Regan2022,Zheng2023,RuizTijerina2020}.

\section{Experimental Section}
\subsection{CL measurements}

CL measurements were performed in a TESCAN MIRA3 instrument equipped with a Schottky field-emission gun and a SPARC Spectral detection system (Delmic). 
Acceleration voltages between 5 and 30\,kV were used with a beam current of 1.4\,nA, measured using a Faraday cup. The CL signal was collected by an aluminium-coated parabolic mirror with a central aperture for the electron beam. 
The mirror provides an effective numerical aperture of 0.97 for collecting generated CL signals. 
The collected CL was directed to a spectrograph (Andor Kymera 193i) equipped with a 150~lines/mm (500\,nm blaze) grating and detected using a CCD camera (Andor Newton). Spectra were acquired with exposure times of 90--120\,s using a beam spot size of approximately 10\,nm to ensure adequate signal-to-noise ratio. Angle-resolved maps were recorded with 90\,s exposure times. All CL spectra and angular maps were background-corrected using dark counts acquired with the electron beam blanked and afterward corrected for the system response function of the instrument~\cite{Ebel2025}. To verify that the observed spectral features are reproducible and not significantly affected by beam-induced modifications, we performed sequential CL measurements, position-dependent CL measurements, and before/after high-contrast secondary-electron imaging on the 100\,nm WS$_2$ flake. These control measurements are discussed in Supplementary Note~S6.

\subsection{Sample fabrication}
The investigated vdW materials were purchased from HQ Graphene. Samples were prepared by mechanical exfoliation and transferred onto polydimethylsiloxane (PDMS) stamps, enabling deterministic placement using a viscoelastic stamping technique~\cite{CastellanosGomez2014}. We utilized TEM windows in a silicon nitride (Si$_3$N$_4$) covered silicon (Si) wafers as target substrates, allowing the selected vdW crystals to be freely suspended while supported at the window edges. Si$_3$N$_4$ film over the windows was removed to enable the free suspension of vdW crystals. 
For deterministic transfer, the substrate was heated to 110$^\circ$C during contact to promote release of the flake from the PDMS stamp. The thickness of the fabricated suspended vdW crystals was measured with optical reflection spectroscopy (For details see Supplementary Note~S2).

\subsection{Dielectric function of a model semiconductor}
We treat the causal optical response of a model semiconductor by calculating the complex dielectric function~\cite{Adachi1990}
\begin{equation}\label{semiconductor}
\begin{aligned}
\epsilon(\omega)=&\frac{2 D}{\pi}  {\left[-\frac{\left(E_g\right)^2}{(\hbar \omega+i \Gamma)^2} \ln \left(\frac{E_c}{E_g}\right)\right.} \\
& +\frac{1}{2}\left(1+\frac{E_g}{\hbar \omega+i \Gamma}\right)^2 \ln \left(\frac{\hbar \omega+i \Gamma+E_c}{\hbar \omega+i \Gamma+E_g}\right) \\
& \left.+\frac{1}{2}\left(1-\frac{E_g}{\hbar \omega+i \Gamma}\right)^2 \ln \left(\frac{\hbar \omega+i \Gamma-E_c}{\hbar \omega+i \Gamma-E_g}\right)\right],
\end{aligned}
\end{equation}
where $E_g$ is the bandgap energy and $E_c$ the cut-off energy. Furthermore, $\Gamma$ corresponds to the lifetime broadening and $D$ represents an overall strength that is independent of the photon energy. Calculations in Fig.~\ref{TR_model} are based on $E_g=1.5$\,eV, $E_c=6$\,eV, $\Gamma=40$\,meV, and $D=60$.

\subsection{Coupled-oscillator model}

To discuss our analytical and theoretical results in the context of photon-exciton hybridization we apply a minimal coupled-oscillator model (see e.g.~\cite{Torma:2015,Tserkezis:2020}),
\begin{gather}
  \begin{bmatrix}
   E_{\mathrm{ph}} & \hbar\Omega_\mathrm{A}/2 &\hbar\Omega_\mathrm{B}/2\\
     \hbar\Omega_\mathrm{A}/2 & E_\mathrm{A} & 0 \\
    \hbar\Omega_\mathrm{B}/2 & 0 &E_\mathrm{B}
   \end{bmatrix}
   \begin{bmatrix} \alpha \\ \beta \\ \gamma \end{bmatrix}
   =\hbar \omega \begin{bmatrix} \alpha \\ \beta \\ \gamma  \end{bmatrix},
\end{gather}
for the resonance energies $\hbar\omega$ of the model. The eigenvectors describe the relative weighting of the excitonic and photonic components. 
Furthermore, $E_j$ and $\hbar\Omega_j$ denote the exciton energies and the associated coupling strength of the $j=\{\mathrm{A},\mathrm{B}\}$ excitons of WS$_2$.  
For a dispersive material $\epsilon(\omega)$, we obtain a frequency-dependent photon energy $E_\mathrm{ph} = E_\mathrm{ph}(\omega)$ through Eq.~\eqref{Dispersion}. Solving the eigenvalue problem for a dispersive material consequently leads to a nonlinear equation
\begin{align}
    E_\mathrm{ph}(\omega) - \hbar \omega - \frac{1}{4}\sum_j\frac{\left(\hbar\Omega_j\right)^2}{E_j-\hbar\omega}=0.
\end{align}
For the generally complex resonant energies obtained as solutions to this equation, we require a model of the permittivity of WS$_2$ that is analytically continued to complex-valued energies. Such a model is obtained by a least-squares fitting of the experimental permittivity of WS$_2$ reported by Munkhbat \emph{et~al.}~\cite{Munkhbat2022}, using in practice a sum of six Lorentzian oscillators (see Supplementary Note~S7). The solutions $\omega$ to this nonlinear equation give the white solid lines in Fig.~\ref{TunableTheo}. For simplicity, the analysis assumed a fixed polar angle of $\theta=60^\circ$.\\

\noindent
\textbf{Competing interests:}
The authors declare no competing financial or non-financial interests.

\noindent
\textbf{Funding Declaration:} The Center for Polariton-driven Light-Matter Interactions (POLIMA) is sponsored by the Danish National Research Foundation (Project No.~DNRF165).

\noindent
\textbf{Author contributions:} The idea of the project was conceived by S.~E and S.~M. 
Sample preparation and CL spectroscopy measurements were handled by S.~E. 
Optical sample characterization were performed by S.~E and M.~N.
The coupled-oscillator model was solved by C.~N.~H. and S.~E.
The project was supervised by S.~M. and N.~A.~M.
All authors contributed to analysing the data and writing the manuscript. 
All authors have accepted responsibility for the entire content of this manuscript and approved its submission.

\noindent
\textbf{Acknowledgments:} We thank P.~A.~D. Gonçalves and C.~Tserkezis for stimulating discussions.

\bibliography{bibliography}

\begin{thebibliography}{10}

\bibitem{Schneider2018}
C.~Schneider, M.~M. Glazov, T.~Korn, S.~H\"{o}fling, and B.~Urbaszek.
\newblock Two-dimensional semiconductors in the regime of strong light-matter coupling.
\newblock {\em Nature Communications}, 9:2695, 2018.

\bibitem{Weisbuch1992}
C.~Weisbuch, M.~Nishioka, A.~Ishikawa, and Y.~Arakawa.
\newblock Observation of the coupled exciton-photon mode splitting in a semiconductor quantum microcavity.
\newblock {\em Physical Review Letters}, 69(23):3314–3317, 1992.

\bibitem{Zhang2018}
L.~Zhang, R.~Gogna, W.~Burg, E.~Tutuc, and H.~Deng.
\newblock Photonic-crystal exciton-polaritons in monolayer semiconductors.
\newblock {\em Nature Communications}, 9:713, 2018.

\bibitem{Sugawara2006}
Y.~Sugawara, T.~A. Kelf, J.~J. Baumberg, M.~E Abdelsalam, and P.~N. Bartlett.
\newblock Strong coupling between localized plasmons and organic excitons in metal nanovoids.
\newblock {\em Physical Review Letters}, 97(26):266808, 2006.

\bibitem{Liu2016}
W.~Liu, B.~Lee, C.~H. Naylor, H.-S. Ee, J.~Park, A.~T.~C. Johnson, and R.~Agarwal.
\newblock Strong exciton–plasmon coupling in {MoS$_2$} coupled with plasmonic lattice.
\newblock {\em Nano Letters}, 16(2):1262–1269, 2016.

\bibitem{Todisco2020}
F.~Todisco, R.~Malureanu, C.~Wolff, P.~A.~D. Gon\c{c}alves, A.~S. Roberts, N.~A. Mortensen, and C.~Tserkezis.
\newblock Magnetic and electric {Mie}-exciton polaritons in silicon nanodisks.
\newblock {\em Nanophotonics}, 9(4):803–814, 2020.

\bibitem{Wang2020}
S.~Wang, T.~V. Raziman, S.~Murai, G.~W. Castellanos, P.~Bai, A.~M. Berghuis, R.~H. Godiksen, A.~G. Curto, and J.~Gómez~Rivas.
\newblock Collective {Mie} exciton-polaritons in an atomically thin semiconductor.
\newblock {\em Journal of Physical Chemistry C}, 124(35):19196–19203, 2020.

\bibitem{Maggiolini2023}
E.~Maggiolini, L.~Polimeno, F.~Todisco, A.~Di~Renzo, B.~Han, M.~De~Giorgi, V.~Ardizzone, C.~Schneider, R.~Mastria, A.~Cannavale, M.~Pugliese, L.~De~Marco, A.~Rizzo, V.~Maiorano, G.~Gigli, D.~Gerace, D.~Sanvitto, and D.~Ballarini.
\newblock Strongly enhanced light–matter coupling of monolayer {WS$_2$} from a bound state in the continuum.
\newblock {\em Nature Materials}, 22(8):964–969, 2023.

\bibitem{Weber2023}
T.~Weber, L.~K\"{u}hner, L.~Sortino, A.~Ben~Mhenni, N.~P. Wilson, J.~K\"{u}hne, J.~J. Finley, S.~A. Maier, and A.~Tittl.
\newblock Intrinsic strong light-matter coupling with self-hybridized bound states in the continuum in {van der Waals} metasurfaces.
\newblock {\em Nature Materials}, 22(8):970–976, 2023.

\bibitem{Dufferwiel2015}
S.~Dufferwiel, S.~Schwarz, F.~Withers, A.~A.~P. Trichet, F.~Li, M.~Sich, O.~Del Pozo-Zamudio, C.~Clark, A.~Nalitov, D.~D. Solnyshkov, G.~Malpuech, K.~S. Novoselov, J.~M. Smith, M.~S. Skolnick, D.~N. Krizhanovskii, and A.~I. Tartakovskii.
\newblock Exciton–polaritons in {van der Waals} heterostructures embedded in tunable microcavities.
\newblock {\em Nature Communications}, 6:8579, 2015.

\bibitem{Urbonas2016}
D.~Urbonas, T.~St\"{o}ferle, F.~Scafirimuto, U.~Scherf, and R.~F. Mahrt.
\newblock Zero-dimensional organic exciton–polaritons in tunable coupled {Gaussian} defect microcavities at room temperature.
\newblock {\em ACS Photonics}, 3(9):1542–1545, 2016.

\bibitem{Lackner2021}
L.~Lackner, M.~Dusel, O.~A. Egorov, B.~Han, H.~Knopf, F.~Eilenberger, S.~Schr\"{o}der, K.~Watanabe, T.~Taniguchi, S.~Tongay, C.~Anton-Solanas, S.~H\"{o}fling, and C.~Schneider.
\newblock Tunable exciton-polaritons emerging from {WS$_2$} monolayer excitons in a photonic lattice at room temperature.
\newblock {\em Nature Communications}, 12:4933, 2021.

\bibitem{VuCam2025}
N.~Vu~Cam, M.~A. Rahman, S.~Akhil, E.~G. Durmusoglu, T.~T.~H. Do, P.~L. Hernandez-Martinez, C.~Dabard, D.~Arora, S.~Uddin, G.~Zamiri, H.~Wang, S.~T. Ha, N.~A. Mortensen, H.~V. Demir, and J.~K.~W. Yang.
\newblock Tunable multiresonant microcavity exciton-polaritons in colloidal quantum wells.
\newblock {\em Nano Letters}, 25(15):6109–6116, 2025.

\bibitem{Sidler2016}
M.~Sidler, P.~Back, O.~Cotlet, A.~Srivastava, T.~Fink, M.~Kroner, E.~Demler, and A.~Imamoglu.
\newblock Fermi polaron-polaritons in charge-tunable atomically thin semiconductors.
\newblock {\em Nature Physics}, 13(3):255–261, 2016.

\bibitem{Munkhbat2020}
B.~Munkhbat, D.~G. Baranov, A.~Bisht, M.~A. Hoque, B.~Karpiak, S.~P. Dash, and T.~Shegai.
\newblock Electrical control of hybrid monolayer tungsten disulfide–plasmonic nanoantenna light–matter states at cryogenic and room temperatures.
\newblock {\em ACS Nano}, 14(1):1196–1206, 2020.

\bibitem{Hoekstra2026}
T.~Hoekstra and J.~van~de Groep.
\newblock Electrically tunable strong coupling in a hybrid-{2D} excitonic metasurface for optical modulation.
\newblock {\em Light: Science \& Applications}, 15:28, 2026.

\bibitem{Ahn2020}
W.~Ahn and B.~S. Simpkins.
\newblock Spectroelectrochemical measurement and modulation of exciton-polaritons.
\newblock {\em APL Photonics}, 5(7):076107, 2020.

\bibitem{Friedrich2025}
D.~Friedrich, J.~Qin, B.~Schurr, T.~Tufarelli, H.~Groß, and B.~Hecht.
\newblock Anticrossing of a plasmonic nanoresonator mode and a single quantum dot at room temperature.
\newblock {\em Advanced Science}, 12(36):e06676, 2025.

\bibitem{Rodriguez2013}
S.~R.~K. Rodriguez and J.~Gómez~Rivas.
\newblock Surface lattice resonances strongly coupled to rhodamine {6G} excitons: tuning the plasmon-exciton-polariton mass and composition.
\newblock {\em Optics Express}, 21(22):27411, 2013.

\bibitem{Guo2017}
R.~Guo, T.~K. Hakala, and P.~T\"orm\"a.
\newblock Geometry dependence of surface lattice resonances in plasmonic nanoparticle arrays.
\newblock {\em Physical Review B}, 95(15):155423, 2017.

\bibitem{Canales2021}
A.~Canales, D.~G. Baranov, T.~J. Antosiewicz, and T.~Shegai.
\newblock Abundance of cavity-free polaritonic states in resonant materials and nanostructures.
\newblock {\em Journal of Chemical Physics}, 154(2):024701, 2021.

\bibitem{vanVugt2011}
L.~K. van Vugt, B.~Piccione, C.-H. Cho, P.~Nukala, and R.~Agarwal.
\newblock One-dimensional polaritons with size-tunable and enhanced coupling strengths in semiconductor nanowires.
\newblock {\em Proceedings of the National Academy of Sciences}, 108(25):10050–10055, 2011.

\bibitem{Verre2019}
R.~Verre, D.~G. Baranov, B.~Munkhbat, J.~Cuadra, M.~K\"{a}ll, and T.~Shegai.
\newblock Transition metal dichalcogenide nanodisks as high-index dielectric {Mie} nanoresonators.
\newblock {\em Nature Nanotechnology}, 14(7):679–683, 2019.

\bibitem{Tserkezis2024}
C.~Tserkezis, P.~E. Stamatopoulou, C.~Wolff, and N.~A. Mortensen.
\newblock Self-hybridisation between interband transitions and {Mie} modes in dielectric nanoparticles.
\newblock {\em Nanophotonics}, 13(14):2513–2522, 2024.

\bibitem{Wang2016}
Q.~Wang, L.~Sun, B.~Zhang, C.~Chen, X.~Shen, and W.~Lu.
\newblock Direct observation of strong light-exciton coupling in thin {WS$_2$} flakes.
\newblock {\em Optics Express}, 24(7):7151, 2016.

\bibitem{Munkhbat2018}
B.~Munkhbat, D.~G. Baranov, M.~St\"{u}hrenberg, M.~Wers\"{a}ll, A.~Bisht, and T.~Shegai.
\newblock Self-hybridized exciton-polaritons in multilayers of transition metal dichalcogenides for efficient light absorption.
\newblock {\em ACS Photonics}, 6(1):139–147, 2018.

\bibitem{Dirnberger2023}
F.~Dirnberger, J.~Quan, R.~Bushati, G.~M. Diederich, M.~Florian, J.~Klein, K.~Mosina, Z.~Sofer, X.~Xu, A.~Kamra, F.~J. García-Vidal, A.~Alù, and V.~M. Menon.
\newblock Magneto-optics in a {van der Waals} magnet tuned by self-hybridized polaritons.
\newblock {\em Nature}, 620(7974):533–537, 2023.

\bibitem{Black2024}
M.~Black, M.~Asadi, P.~Darman, S.~Se\c{c}kin, F.~Schillm\"{o}ller, T.~A.~F. K\"{o}nig, S.~Darbari, and N.~Talebi.
\newblock Long-range self-hybridized exciton-polaritons in two-dimensional {Ruddlesden–Popper} perovskites.
\newblock {\em ACS Photonics}, 11(10):4065--4075, 2024.

\bibitem{Anantharaman2024}
S.~B. Anantharaman, J.~Lynch, C.~E. Stevens, C.~Munley, C.~Li, J.~Hou, H.~Zhang, A.~Torma, T.~Darlington, F.~Coen, K.~Li, A.~Majumdar, P.~J. Schuck, A.~Mohite, H.~Harutyunyan, J.~R. Hendrickson, and D.~Jariwala.
\newblock Dynamics of self-hybridized exciton–polaritons in {2D} halide perovskites.
\newblock {\em Light: Science \& Applications}, 13:1, 2024.

\bibitem{Ziegler2025}
J.~D. Ziegler, S.~Papadopoulos, A.~J. Moilanen, M.~Martínez, Q.~Lin, K.~Mosina, T.~Taniguchi, K.~Watanabe, Z.~Sofer, F.~Dirnberger, and L.~Novotny.
\newblock Electrical excitation of self-hybridized exciton polaritons in a {van der Waals} antiferromagnet.
\newblock {\em Science Advances}, 11(45):eadz6724, 2025.

\bibitem{Kociak_2025}
M.~Kociak, X.~Li, Y.~Auad, C.~Hamon, C.~Goldmann, L.~Tizei, and O.~Stéphan.
\newblock Strong and weak coupling nanophysics with free electron beams.
\newblock {\em Advanced Optical Materials}, 13(34):e01548, 2025.

\bibitem{Song2021}
J.-H. Song, S.~Raza, J.~van~de Groep, J.-H. Kang, Q.~Li, P.~G. Kik, and M.~L. Brongersma.
\newblock Nanoelectromechanical modulation of a strongly-coupled plasmonic dimer.
\newblock {\em Nature Communications}, 12:48, 2021.

\bibitem{Yankovich2019}
A.~B. Yankovich, B.~Munkhbat, D.~G. Baranov, J.~Cuadra, E.~Olsén, H.~Louren\c{c}o-Martins, L.~H.~G. Tizei, M.~Kociak, E.~Olsson, and T.~Shegai.
\newblock Visualizing spatial variations of plasmon–exciton polaritons at the nanoscale using electron microscopy.
\newblock {\em Nano Letters}, 19(11):8171–8181, 2019.

\bibitem{Bitton2020}
O.~Bitton, S.~N. Gupta, L.~Houben, M.~Kvapil, V.~Křápek, T.~Šikola, and G.~Haran.
\newblock Vacuum {Rabi} splitting of a dark plasmonic cavity mode revealed by fast electrons.
\newblock {\em Nature Communications}, 11:487, 2020.

\bibitem{Chahshouri2022}
F.~Chahshouri, M.~Taleb, F.~K Diekmann, K~Rossnagel, and N.~Talebi.
\newblock Interaction of excitons with {Cherenkov} radiation in {WSe$_2$} beyond the non-recoil approximation.
\newblock {\em Journal of Physics D: Applied Physics}, 55(14):145101, 2022.

\bibitem{Nerl2024}
H.~C. Nerl, K.~Elyas, Z.~Kochovski, N.~Talebi, C.~T. Koch, and K.~H\"{o}flich.
\newblock Flat dispersion at large momentum transfer at the onset of exciton polariton formation.
\newblock {\em Communications Physics}, 7:388, 2024.

\bibitem{Davoodi2022}
F.~Davoodi, M.~Taleb, F.~K. Diekmann, T.~Coenen, K.~Rossnagel, and N.~Talebi.
\newblock Tailoring the band structure of plexcitonic crystals by strong coupling.
\newblock {\em ACS Photonics}, 9(7):2473–2482, 2022.

\bibitem{Taleb2021}
M.~Taleb, F.~Davoodi, F.~K. Diekmann, K.~Rossnagel, and N.~Talebi.
\newblock Charting the exciton–polariton landscape of {WSe$_2$} thin flakes by cathodoluminescence spectroscopy.
\newblock {\em Advanced Photonics Research}, 3(1):2100124, 2021.

\bibitem{Ginzburg1996}
V.~L. Ginzburg.
\newblock Radiation by uniformly moving sources ({Vavilov--Cherenkov} effect, transition radiation, and other phenomena).
\newblock {\em Physics-Uspekhi}, 39(10):973–982, 1996.

\bibitem{Ritchie1962}
R.~H. Ritchie and H.~B. Eldridge.
\newblock Optical emission from irradiated foils. {I}.
\newblock {\em Physical Review}, 126(6):1935--1947, 1962.

\bibitem{Arakawa1964}
E.~T. Arakawa, N.~O. Davis, and R.~D. Birkhoff.
\newblock Temperature and thickness dependence of transition radiation from thin silver foils.
\newblock {\em Physical Review}, 135(1A):A224--A226, 1964.

\bibitem{Yamamoto1996}
N.~Yamamoto, H.~Sugiyama, and A.~Toda.
\newblock Cherenkov and transition radiation from thin plate crystals detected in the transmission electron microscope.
\newblock {\em Proceedings of the Royal Society A}, 452(1953):2279--2301, 1996.

\bibitem{Lin2018}
X.~Lin, S.~Easo, Y.~Shen, H.~Chen, B.~Zhang, J.~D. Joannopoulos, M.~Soljačić, and I.~Kaminer.
\newblock Controlling cherenkov angles with resonance transition radiation.
\newblock {\em Nature Physics}, 14(8):816–821, 2018.

\bibitem{Fiedler2022}
S.~Fiedler, P.~E. Stamatopoulou, A.~Assadillayev, C.~Wolff, H.~Sugimoto, M.~Fujii, N.~A. Mortensen, S.~Raza, and C.~Tserkezis.
\newblock Disentangling cathodoluminescence spectra in nanophotonics: Particle eigenmodes vs transition radiation.
\newblock {\em Nano Letters}, 22(6):2320–2327, 2022.

\bibitem{Munkhbat2022}
B.~Munkhbat, P.~Wróbel, T.~J. Antosiewicz, and T.~O. Shegai.
\newblock Optical constants of several multilayer transition metal dichalcogenides measured by spectroscopic ellipsometry in the 300–1700 nm range: High index, anisotropy, and hyperbolicity.
\newblock {\em ACS Photonics}, 9(7):2398–2407, 2022.

\bibitem{Hill2018}
H.~M. Hill, A.~F. Rigosi, S.~Krylyuk, J.~Tian, N.~V. Nguyen, A.~V. Davydov, D.~B. Newell, and A.~R. Hight~Walker.
\newblock Comprehensive optical characterization of atomically thin {NbSe$_2$}.
\newblock {\em Physical Review B}, 98(16):165109, 2018.

\bibitem{Ebel2025}
S.~Ebel, Y.~Lebsir, T.~Yezekyan, N.~A. Mortensen, and S.~Morozov.
\newblock An atlas of photonic and plasmonic materials for cathodoluminescence microscopy.
\newblock {\em Nanophotonics}, 14(15):2647--2671, 2025.

\bibitem{Nrgaard2025}
M.~Nørgaard, T.~Yezekyan, S.~Rolfs, C.~Frydendahl, N.~A. Mortensen, and V.~A. Zenin.
\newblock Near-field refractometry of {van der Waals} crystals.
\newblock {\em Nanophotonics}, 14(14):2473–2483, 2025.

\bibitem{Ermolaev2021}
G.~A. Ermolaev, D.~V. Grudinin, Y.~V. Stebunov, K.~V. Voronin, V.~G. Kravets, J.~Duan, A.~B. Mazitov, G.~I. Tselikov, A.~Bylinkin, D.~I. Yakubovsky, S.~M. Novikov, D.~G. Baranov, A.~Y. Nikitin, I.~A. Kruglov, T.~Shegai, P.~Alonso-González, A.~N. Grigorenko, A.~V. Arsenin, K.~S. Novoselov, and V.~S. Volkov.
\newblock Giant optical anisotropy in transition metal dichalcogenides for next-generation photonics.
\newblock {\em Nature Communications}, 12:854, 2021.

\bibitem{DiGiulio2024}
V.~Di~Giulio, E.~Akerboom, A.~Polman, and F.~J. García~de Abajo.
\newblock Toward optimum coupling between free electrons and confined optical modes.
\newblock {\em ACS Nano}, 18(22):14255--14275, 2024.

\bibitem{Ai2025}
R.~Ai, X.~Cui, Y.~Li, and X.~Zhuo.
\newblock Local strain engineering of two-dimensional transition metal dichalcogenides towards quantum emitters.
\newblock {\em Nano-Micro Letters}, 17(1), January 2025.

\bibitem{Frg2019}
M.~F\"{o}rg, L.~Colombier, R.~K. Patel, J.~Lindlau, A.~D. Mohite, H.~Yamaguchi, M.~M. Glazov, D.~Hunger, and A.~H\"{o}gele.
\newblock Cavity-control of interlayer excitons in {van der Waals} heterostructures.
\newblock {\em Nature Communications}, 10:3697, 2019.

\bibitem{Moore2025}
S.~L. Moore, H.~Y. Lee, N.~Rivera, Y.~Karube, M.~Ziffer, E.~S. Yanev, T.~P. Darlington, A.~J. Sternbach, M.~A. Holbrook, J.~Pack, X.~Xu, C.~R. Dean, J.~S. Owen, P.~J. Schuck, M.~Delor, X.~Y. Zhu, J.~Hone, and D.~N. Basov.
\newblock {Van der Waals} waveguide quantum electrodynamics probed by infrared nano-photoluminescence.
\newblock {\em Nature Photonics}, 19(8):833–839, 2025.

\bibitem{Regan2022}
E.~C. Regan, D.~Wang, E.~Y. Paik, Y.~Zeng, L.~Zhang, J.~Zhu, A.~H. MacDonald, H.~Deng, and F.~Wang.
\newblock Emerging exciton physics in transition metal dichalcogenide heterobilayers.
\newblock {\em Nature Reviews Materials}, 7(10):778–795, 2022.

\bibitem{Zheng2023}
H.~Zheng, B.~Wu, S.~Li, J.~Ding, J.~He, Z.~Liu, C.-T. Wang, J.-T. Wang, A.~Pan, and Y.~Liu.
\newblock Localization-enhanced moiré exciton in twisted transition metal dichalcogenide heterotrilayer superlattices.
\newblock {\em Light: Science \& Applications}, 12:117, 2023.

\bibitem{RuizTijerina2020}
D.~A. Ruiz-Tijerina, I.~Soltero, and F.~Mireles.
\newblock Theory of moiré localized excitons in transition metal dichalcogenide heterobilayers.
\newblock {\em Physical Review B}, 102(19):95403, 2020.

\bibitem{CastellanosGomez2014}
A.~Castellanos-Gomez, M.~Buscema, R.~Molenaar, V.~Singh, L.~Janssen, H.~S.~J. van~der Zant, and G.~A. Steele.
\newblock Deterministic transfer of two-dimensional materials by all-dry viscoelastic stamping.
\newblock {\em 2D Materials}, 1(1):011002, 2014.

\bibitem{Adachi1990}
S.~Adachi.
\newblock Effects of the indirect transitions on optical dispersion relations.
\newblock {\em Physical Review B}, 41(6):3504--3508, 1990.

\bibitem{Torma:2015}
P.~Törmä and W.~L. Barnes.
\newblock Strong coupling between surface plasmon polaritons and emitters: a review.
\newblock {\em Reports on Progress in Physics}, 78(1):013901, 2015.

\bibitem{Tserkezis:2020}
C.~Tserkezis, A.~I. Fernández-Domínguez, P.~A.~D. Gonçalves, F.~Todisco, J.~D. Cox, K.~Busch, N.~Stenger, S.~I. Bozhevolnyi, N.~A. Mortensen, and C.~Wolff.
\newblock On the applicability of quantum-optical concepts in strong-coupling nanophotonics.
\newblock {\em Reports on Progress in Physics}, 83(8):082401, 2020.

\end{thebibliography}
\bibliographystyle{unsrt}



\end{document}